\documentstyle[epsfig]{aipproc}

\newcommand{\rgn}{($\gamma$,n)}
\newcommand{\rng}{(n,$\gamma$)}
\newcommand{\rga}{($\gamma$,$\alpha$)}
\newcommand{\rag}{($\alpha$,$\gamma$)}
\newcommand{\rgp}{($\gamma$,p)}

\begin{document}
\title{
  Photon-induced Reactions \\ in Stars and in the Laboratory: \\ 
  A Critical Comparison
}

\author{
  Peter Mohr
}

\address{
  Strahlentherapie, Diakoniekrankenhaus, D-74523 Schw\"abisch Hall, Germany
}

\maketitle

\begin{abstract}
Photon-induced reactions during the astrophysical $p$- (or $\gamma$-)
process occur at typical temperatures of $1.8 \le T_9 \le
3.3$. Experimental data of \rgn , \rgp , or \rga\ reactions -- if
available in the relevant energy region -- cannot be used directly to
measure astrophysical \rgn , \rgp , or \rga\ reaction rates because of
the thermal excitation of target nuclei at these high
temperatures. Usually, statistical model calculations are used to
predict photon-induced reaction rates. The relations between
experimental reaction cross sections, theoretical predictions, and
astrophysical reaction rates will be critically discussed.
\end{abstract}

\section{Introduction}
\label{sec:intro}
The nucleosynthesis of heavy neutron-deficient nuclei, so-called
$p$-nuclei, proceeds mainly via a series of photon-induced \rgn , \rgp
, and \rga\ reactions in the thermal photon bath of an explosive
astrophysical event. Type II supernovae are good candidates to provide
the required astrophysical environment (e.g., temperatures of $1.8 \le
T_9 \le 3.3$) \cite{Arn03,Rau02,Lam92,Woo78}.

Calculations of the astrophysical reaction rates and cross sections
are based on the statistical model; input parameters for
photon-induced reactions are $\gamma$-ray strength functions, optical
potentials, and level densities. Recent results are summarized in
\cite{Rau00,Rau01,Gor03}.

Experimental data for photon-induced cross sections in the
astrophysically relevant energy region have been obtained
using two different techniques. Monochromatic photons from Compton
backscattering of a Laser beam were used by \cite{Uts03}, and a
quasi-thermal photon spectrum was obtained by a superposition of
bremsstrahlung spectra \cite{Mohr00,Vogt01,Vogt02}.

In this paper the astrophysically relevant energy window for \rgn ,
\rgp , and \rga\ reactions \cite{Mohr01,Mohr03} will be analyzed
taking into account that the target nuclei may be thermally excited at
the typical temperatures of $1.8 \le T_9 \le 3.3$. A critical
comparison between experimental data in the laboratory and data for
thermally excited nuclei in stars will be given, and relevant input
parameters for the statistical model will be clearly defined.

\section{Gamow Window for \rgn\ and \rga\ reactions}
\label{sec:gamow}
For simplicity, the following discussion will be restricted to \rgn\
and \rga\ reactions. \rgp\ reactions play only a minor role in the
reaction network for $p$-process nucleosynthesis. Additionally, most
of the arguments given for \rga\ reactions are valid for \rgp\
reactions, too.

The nucleus $^{148}$Gd and the reactions $^{148}$Gd\rgn $^{147}$Gd and
$^{148}$Gd\rga $^{144}$Sm will be chosen as an example because the
nucleosynthesis path of the $p$-process shows a branching point
between \rgn\ and \rga\ reactions which defines the production
ratio between $^{146}$Sm and $^{144}$Sm. This ratio may be used as a
chronometer for the $p$-process \cite{Arn03,Woo78,Rau95} because it can be
measured at the time of the formation of the solar system from
correlations between the $^{144}$Sm abundance and isotopic anomalies
in $^{142,144}$Nd in meteorites \cite{Pri89}. 

The astrophysical reaction rate $\lambda(T)$ of a
photon-induced reaction is given by
\begin{equation}
\lambda(T) =
  \int_0^\infty 
  c \,\, n_\gamma(E,T) \,\, \sigma_{(\gamma,{\rm{x}})}(E) \,\, dE
\label{eq:gamow}
\end{equation}
with the speed of light $c$ and the cross section of the
$\gamma$-induced reaction $\sigma_{(\gamma,{\rm{x}})}(E)$. The thermal
photon density $n_\gamma(E,T)$ is given by the Planck distribution
\begin{equation}
n_\gamma(E,T) = 
  \left( \frac{1}{\pi} \right)^2 \,
  \left( \frac{1}{\hbar c} \right)^3 \,
  \frac{E^2}{\exp{(E/kT)} - 1}
\label{eq:planck}
\end{equation}
where $n_\gamma(E,T)$ is the number of $\gamma$-rays at energy $E$ per
unit of volume and energy interval. The integrand in Eq.~(\ref{eq:gamow})
is defined by the product of the cross section which increases with
energy and the photon density which decreases exponentially with
energy. This leads to a well-defined energy window which is
astrophysically relevant (the so-called Gamow window). A comparison of
typical Gamow windows for \rgn\ and \rga\ reactions for target nuclei
in their ground states is given in \cite{Mohr03}.


\subsection{$^{148}$Gd\rgn $^{147}$Gd}
\label{sec:rgn}
\begin{figure}[hbt]
\includegraphics[ bb = 20 45 555 375, width = 14.0cm, clip]{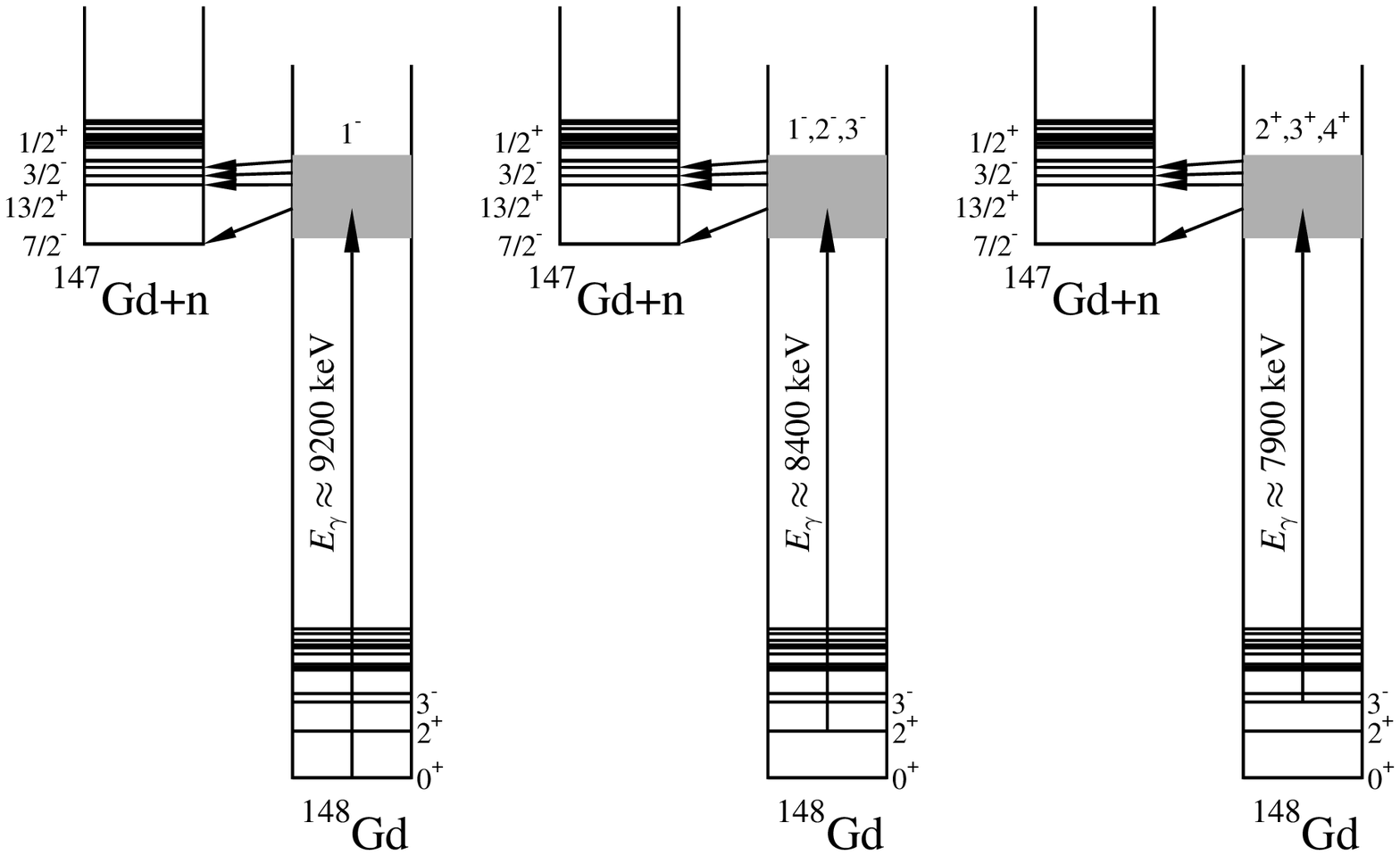}
\caption{
\label{fig:gamow_gn}
Gamow window for the $^{148}$Gd\rgn $^{147}$Gd reaction for the ground
state of $^{148}$Gd ($0^+, 0\,{\rm{keV}}$; left) and the first excited
states ($2^+, 784\,{\rm{keV}}$; middle; and $3^-, 1273\,{\rm{keV}}$;
right). Discussion see Sect.~\ref{sec:rgn}. All level data are from
\protect\cite{ENSDF}. 
}
\end{figure}

The Gamow window for \rgn\ reactions is located close above the
neutron threshold. The maximum of the integrand in
Eq.~(\ref{eq:gamow}) is located at $E_0^{\rm{n}} \approx S_{\rm{n}} +
kT/2 \approx 9200$\,keV for $T_9 = 2.5$ where $S_{\rm{n}}$ is the
neutron separation energy $S_{\rm{n}}(^{148}{\rm{Gd}}) =
8984$\,keV. The typical width of this window is about
1\,MeV. Therefore, the astrophysically relevant window for the
excitation energy $E_x$ is located between $S_{\rm{n}}$ and $S_{\rm{n}} +
1$\,MeV (see Fig.~\ref{fig:gamow_gn}). The position of the Gamow
window for \rgn\ reactions depends only weakly on the temperature.

If the nucleus $^{148}$Gd is in its $0^+$ ground state, the dominating
$E1$ transitions lead to $1^-$ states in the Gamow window (left part
of Fig.~\ref{fig:gamow_gn}, gray shaded area). These $1^-$ states may
decay by neutron emission to low-lying states in $^{147}$Gd with
$E_x(^{147}{\rm{Gd}}) < 1$\,MeV. Note that there is no Coulomb barrier
for neutrons, and because of the small centrifugal barrier transitions
to states with low $J^\pi$ in $^{147}$Gd are preferred. The cross
section for this process can be measured in the laboratory. A
statistical model prediction of this cross section requires the $E1$
$\gamma$-ray strength function around the energy $E_0^{\rm{n}}$ for
the excitation process. Neutron and $\alpha$ optical potentials, the
$\gamma$-ray strength function at $E < E_0^{\rm{n}}$, and the level
density of the residual nuclei above experimentally known levels are
required for the calculation of the possible decays of excited
$^{148}$Gd by neutron, $\alpha$, or $\gamma$ emission.

The situation changes if the nucleus $^{148}$Gd is not in the ground
state, but thermally excited to its low-lying levels. For simplicity,
the discussion is restricted to the first two levels at $E_x =
784$\,keV ($2^+$) and 1273\,keV ($3^-$). A significant contribution of
these levels is already obtained at temperatures $kT < E_x$ because
the ratio of population $n_x/n_0$ is given by the Boltzmann factor
$\exp{(-E_x/kT)}$ and by the statistical weight of the spins
\begin{equation}
\frac{n_x}{n_0} = \frac{2J_x+1}{2J_0+1} \, \exp{(-E_x/kT)} = (2J_x+1)
\, \exp{(-E_x/kT)}
\label{eq:ratio}
\end{equation}
for even-even nuclei with $J^\pi_0 = 0^+$.  Assuming a similar energy
dependence of the \rgn\ cross section of the excited state, one finds
again a Gamow window close above the threshold at excitation energies
around $E_0^{\rm{n}}$. However, the required photon energy for a \rgn\
reaction is reduced by the excitation energy of the populated
low-lying state: $E_\gamma = E_0^{\rm{n}} - E_x(2^+) \approx
8400$\,keV and $E_\gamma = E_0^{\rm{n}} - E_x(3^-) \approx
7900$\,keV. Starting from the $2^+$ ($3^-$) state, $E1$ transitions
may populate states with $J^\pi = 1^-, 2^-, 3^-$ ($J^\pi = 2^+, 3^+,
4^+$) as shown in Fig.~\ref{fig:gamow_gn}, middle and right.  These
states may decay by neutron emission to low-lying states in
$^{147}$Gd, again preferring final states with small spin
differences. This process cannot be measured in the laboratory. A
statistical model calculation for these processes starting from the
thermally excited $2^+$ ($3^-$) state requires the $E1$ $\gamma$-ray
strength function around the energy $E_\gamma = E_0^{\rm{n}} -
E_x(2^+) \approx 8400$\,keV resp.\ $E_\gamma = E_0^{\rm{n}} - E_x(3^-)
\approx 7900$\,keV for the excitation. For the decay the same
ingredients as in the previous case are required.

The important results for the $^{148}$Gd\rgn $^{147}$Gd reaction are
that ($i$) excitation energies around $E_0^{\rm{n}} \approx 9200$\,keV
are the relevant region independent of the thermal excitation of
$^{148}$Gd, and ($ii$) the $E1$ $\gamma$-ray strength function has to
be known at the energy $E_0^{\rm{n}} \approx 9200$\,keV for $^{148}$Gd
in the ground state and at lower energies $E_0^{\rm{n}} - E_x$ for
$^{148}$Gd in thermally excited states. Note that the $E1$
$\gamma$-ray strength function at energies $E_0^{\rm{n}} - E_x <
S_{\rm{n}}$ cannot be measured by \rgn\ reactions because this
strength is located below threshold! A similar phenomenon of important
$\gamma$-ray strength below threshold has been found for neutron
capture cross sections relevant for the $r$-process
\cite{Gor98}. Usually, one extrapolates the $E1$ $\gamma$-ray strength
function from the giant dipole resonance (GDR) to lower energies, and
one assumes, following the Brink-Axel hypothesis
\cite{Axe61,Ros68,Bar73}, that a similar GDR and $E1$ $\gamma$-ray
strength distribution can be found above each excited
state. Such extrapolations of the $\gamma$-ray strength function
towards lower energies are extensively discussed in \cite{Kop90}.

\subsection{$^{148}$Gd\rga $^{144}$Sm}
\label{sec:rga}
The position of the Gamow window for \rga\ reactions is mainly defined
by the Coulomb barrier. The maximum of the integrand in
Eq.~(\ref{eq:gamow}) for \rga\ reactions is shifted by the $\alpha$
separation energy $S_\alpha$ compared to the \rag\ reaction:
\begin{equation}
E_0^\alpha = 1.22\,(Z_P^2 \, Z_T^2 A_{\rm{red}} \, T_6^2)^{1/3}\,{\rm{keV}} +
S_\alpha
\label{eq:gamow_ga}
\end{equation}
The Gamow window for \rga\ reactions is much broader compared to the
\rgn\ reaction, and because many heavy neutron-deficient nuclei are
$\alpha$ unbound ($S_\alpha < 0$) the energy $E_0^\alpha$ is often
smaller than $E_0^{\rm{n}}$. The position of the Gamow window depends
sensitively on the temperature $T$. For $T_9 = 2.5$ one finds the
Gamow window at $E_0^\alpha = 5520$\,keV (with $S_\alpha =
-3271$\,keV) and with a width of about 3180\,keV. $T_9 = 2.0$ ($3.0$)
leads to $E_0^\alpha = 4300$\,keV ($6660$\,keV).

\begin{figure}[hbt]
\includegraphics[ bb = 20 112 555 422, width = 14.0cm, clip]{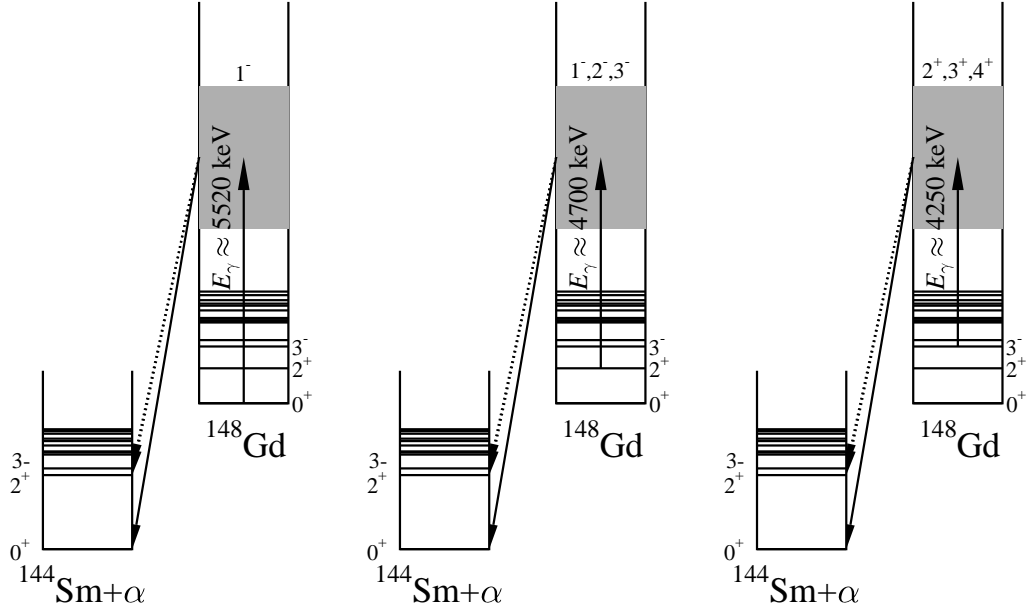}
\caption{
\label{fig:gamow_ga}
Gamow window for the $^{148}$Gd\rga $^{144}$Sm reaction for the ground
state of $^{148}$Gd ($0^+; 0\,{\rm{keV}}$; left) and the first excited
states ($2^+; 784\,{\rm{keV}}$; middle; and $3^-; 1273\,{\rm{keV}}$;
right) at a temperature of $T_9 = 2.5$. Further discussion see
Sect.~\ref{sec:rga}. All level data are from \protect\cite{ENSDF}.
}
\end{figure}

If the nucleus $^{148}$Gd is in its $0^+$ ground state, the dominating
$E1$ transitions lead to $1^-$ states in the Gamow window (left part
of Fig.~\ref{fig:gamow_ga}, gray shaded area). These $1^-$ states may
decay by $\alpha$ emission to low-lying states in $^{144}$Sm. Because
of the Coulomb barrier for $\alpha$ particles, the decay to the ground
state of $^{144}$Sm will be preferred; transitions to excited states
in $^{144}$Sm are suppressed because of the reduced tunneling
probability; they are shown as dashed lines in
Fig.~\ref{fig:gamow_ga}. The cross section for this process can be
measured in the laboratory. A statistical model prediction of this
cross section requires the $E1$ $\gamma$-ray strength function around
the energy $E_0^{\rm{\alpha}}$ for the excitation. The $\alpha$
optical potential, the $\gamma$-ray strength function at $E <
E_0^{\rm{\alpha}}$, and the level density of the residual nuclei above
experimentally known levels are required for the calculation of the
possible decays of excited $^{148}$Gd by $\alpha$ or $\gamma$
emission; the neutron channel is not open at the low excitation
energies. 

Again, the situation changes if the nucleus $^{148}$Gd is not in the
ground state, but thermally excited to its low-lying levels.  Assuming
a similar energy dependence of the \rga\ cross section of the excited
state, one finds again a Gamow window at excitation energies around
$E_0^{{\alpha}} \approx 5520$\,keV. However, the required photon
energy for a \rga\ reaction is reduced by the excitation energy of the
populated low-lying state: $E_\gamma = E_0^{{\alpha}} - E_x(2^+)
\approx 4700$\,keV and $E_\gamma = E_0^{{\alpha}} - E_x(3^-) \approx
4250$\,keV. Starting from the $2^+$ ($3^-$) state, $E1$ transitions
may populate states with $J^\pi = 1^-, 2^-, 3^-$ ($J^\pi = 2^+, 3^+,
4^+$) as shown in Fig.~\ref{fig:gamow_ga}, middle and right. These
states may decay by $\alpha$ emission to low-lying states in
$^{144}$Sm, again preferring the ground state of $^{144}$Sm because of
the Coulomb barrier (with the exception of the unnatural parity states
with $J^\pi = 2^-$ and $3^+$). This process cannot be measured in the
laboratory. A statistical model calculation for these processes
starting from the thermally excited $2^+$ ($3^-$) state requires the
$E1$ $\gamma$-ray strength function around the energy $E_\gamma =
E_0^{{\alpha}} - E_x(2^+) \approx 4700$\,keV resp.\ $E_\gamma =
E_0^{{\alpha}} - E_x(3^-) \approx 4250$\,keV for the excitation. For
the decay the same ingredients as in the previous case are required.

The important results for the $^{148}$Gd\rga $^{144}$Sm reaction at
$T_9 = 2.5$ are that ($i$) the excitation energies around
$E_0^{{\alpha}} \approx 5520$\,keV are the relevant region independent
of the thermal excitation of $^{148}$Gd (but $E_0^\alpha$ itself
depends sensitively on the temperature!), and ($ii$) the $E1$
$\gamma$-ray strength function has to be known at the energy
$E_0^\alpha$ for $^{148}$Gd in the ground state and at lower energies
$E_0^\alpha - E_x$ for $^{148}$Gd in thermally excited states. Note
that the $E1$ $\gamma$-ray strength function has now to be known at
relatively low energies.

\subsection{The ratio \rgn /\rga }
\label{sec:ratio}
As stated above, the branching ratio between \rgn\ and \rga\ reactions
defines the nucleosynthesis in the $p$-process. For a reliable
prediction of branchings between \rgn\ and \rga\ reactions from
statistical model calculations, various ingredients have to be known
accurately. Besides the optical potentials for neutrons and $\alpha$
particles and the level densities, $E1$ $\gamma$-ray strength
functions have to be known at different energies for the \rgn\ and
\rga\ reactions. Therefore a precise knowledge of the energy
dependence of the $\gamma$-ray strength function at low energies is
essential for the prediction of the ratio of \rgn /\rga\ cross
sections. It is highly desirable to check all ingredients of the model
calculations by experimental data including the Brink-Axel hypothesis
\cite{Axe61,Ros68,Bar73} where a partial breakdown was discussed in
\cite{Kop90}.

\section{Comparison with experimental data}
\label{sec:expdata}
It is found that discrepancies between different statistical model
calculations are mainly caused by different extrapolations of the $E1$
$\gamma$-ray strength function and by different $\alpha$ optical
potentials whereas various parametrizations of level densities and
neutron optical potentials lead to almost identical predictions for
the cross section. The study of a global $\alpha$ optical potential
has been described elsewhere
(see Refs.~\cite{Atz96,Moh97,Moh00a,Ful01,Dem02}). 

The $E1$ $\gamma$-ray strength function was determined for many nuclei
from photoabsorption data around the GDR \cite{Die88}. However, the
astrophysically relevant energy region for \rgn\ and \rga\ reactions
is located at significantly lower energies. \rgn\ data close above the
neutron threshold \cite{Uts03,Mohr00,Vogt01,Vogt02} help to restrict
the $E1$ $\gamma$-ray strength function at energies around
$E_0^{\rm{n}}$. Experimental data with monochromatic photons
\cite{Uts03} should be preferred because such data can be directly
compared to theoretical predictions. The method of the quasi-thermal
photon spectrum using a superposition of bremsstrahlung spectra
\cite{Mohr00} provides averaged cross sections which cannot be
compared to theoretical predictions directly. 

A standard technique to measure $E1$ $\gamma$-ray strength functions
at low energies is photon scattering \cite{Kne96}. Bremsstrahlung
experiments with unpolarized photons have no sufficient sensitivity to
distinguish between $E1$ and $M1$ transitions \cite{Zil02} and are not
well-suited for the precise determination of the $E1$ $\gamma$-ray
strength function and its energy dependence. Off-axis bremssstrahlung
may provide a limited photon polarization. However, the best solution
would be photon scattering experiments using 100\,\% polarized photons
from Laser-Compton scattering. Especially the SPring-8 facility with a
high electron energy of several GeV and a long Laser wavelength of
several $\mu$m provides an almost white spectrum with photon energies
of several MeV, huge intensities, and almost 100\,\% polarization
\cite{Uts03a}. Alternatively, $\gamma$-ray spectra in neutron capture
reactions have been used to extract the $\gamma$-ray strength function
at low energies \cite{Iga86}.

A special problem has to be mentioned. The $\gamma$-ray strength
function is continuous above the neutron threshold but the measured
$E1$ strength below neutron threshold consists of discrete
levels. A direct comparison remains difficult. In the case of \rga\
reactions the Gamow window is typically found at energies below the
neutron threshold. Therefore, the relevant $E1$ strength is again
concentrated in discrete levels, and the \rga\ reaction rate may
depend sensitively on the excitation energies of the corresponding
levels. 

Experimental data for the inverse capture reactions may provide
further insight into the photodisintegration reaction rates. The
reaction $^{144}$Sm\rag $^{148}$Gd was measured close above the
astrophysically relevant energies \cite{Som98}. Under laboratory
conditions the nucleus $^{144}$Sm is in its ground state, and the
\rag\ reaction populates many levels in $^{148}$Gd. These experimental
conditions for the \rag\ reaction are very close to the \rga\ reaction
under stellar conditions. Here many levels of $^{148}$Gd are thermally
populated, and because of tunneling probabilities through the Coulomb
barrier mainly the $^{144}$Sm ground state is populated in the \rga\
reaction (see Fig.~\ref{fig:gamow_ga}). However, laboratory conditions
for the \rga\ reaction with $^{148}$Gd in its ground state differ
significantly from stellar conditions for the \rga\
reaction. Therefore, a measurement of the \rag\ reaction provides the
best test for statistical model predictions of astrophysical \rga\
reaction rates. This argument does not hold for \rgn\ reactions: under
stellar conditions excited states in the target and residual nucleus
have to be taken into account (see Fig.~\ref{fig:gamow_gn}) whereas in
laboratory \rgn\ $[$\rng $]$ experiments the target [residual] nucleus
is in its ground state.

\section{Conclusions}
\label{sec:conc}
Reaction networks for the nucleosynthesis of heavy nuclei require a
huge number of reaction cross sections and reaction rates at high
temperatures which are calculated using the statistical
model. Experimental data are rare in the astrophysically relevant
energy region; additionally, astrophysical reaction rates cannot be
derived directly from experimental data in the laboratory because of
the thermal excitation of target nuclei under stellar
conditions. However, experimental data can provide systematic input
parameters for the statistical model calculations. Improved
$\gamma$-ray strength functions and a global $\alpha$-nucleus
potential are needed.

Although present photodisintegration experiments at astrophysically
relevant energies (e.g., \cite{Uts03,Mohr00,Vogt01,Vogt02}) can
provide valuable information for the theoretical prediction of
reaction rates, there are limitations to the extent of information
because only few relevant transitions are tested experimentally.
Especially for the prediction of astrophysical \rga\ reaction rates, a
measurement of the inverse \rag\ cross section seems to be a better
test for the ingredients of the statistical model than a measurement
of the \rga\ reaction.

The nice idea of producing a quasi-thermal photon spectrum in the
laboratory \cite{Mohr00,Vogt01,Vogt02} is unfortunately in many cases
not really useful because the measured laboratory reaction rates may
differ significantly from the reaction rates at typical stellar
conditions with temperatures of about $1.8 \le T_9 \le 3.3$
\cite{Vogt01}. The relevant ingredients of statistical model
calculations, namely the $E1$ $\gamma$-ray strength function, can be
extracted with improved precision and reliability from experiments
with monochromatic photons \cite{Uts03}.

~\\
Discussions with T.\ Rauscher, H.\ Utsunomiya, and A.\ Zilges are
gratefully acknowledged.

\end{document}